\documentclass[prb, twocolumn, amsmath, amsfonts, amssymb]{revtex4-2}
\usepackage{graphicx}
\usepackage{hyperref}
\usepackage{amsmath}
\usepackage{amsfonts}
\usepackage{amssymb}
\usepackage{float}

\begin{document}

\title{Polarization Transmission in the Electron-Ion Collider's Hadron Storage Ring}
\author{E. Hamwi}
\email{eh652@cornell.edu}
\author{G. H. Hoffstaetter}
\affiliation{Cornell University, Ithaca, NY 14850}

\begin{abstract}
The successful operation of the future Electron-Ion Collider is contingent on maintaining high hadron beam polarization up to 275 GeV. The Hadron Storage Ring lattice, however, features a symmetry-breaking interaction region that excites strong, non-systematic spin resonances, posing a significant threat to polarization preservation. This paper systematically investigates two complementary strategies to ensure high polarization transmission. The first method involves optimizing the vertical betatron phase advance between Siberian snakes to orchestrate a cancellation of depolarizing kicks across the ring. We demonstrate through simulations that both dynamic and fixed-optics solutions based on this principle can successfully preserve polarization through the strongest resonances. The second, more powerful approach involves optimizing the snake rotation axes to suppress resonance driving terms at their source. We revisit established symmetric configurations, such as the Lee-Courant schemes, and introduce a novel, highly symmetric ``Doubly Lee-Courant" (DLC) scheme, which enforces a local \(\pi\) spin phase advance across every consecutive pair of snakes. Our analysis reveals a clear performance hierarchy, with the DLC configuration providing an exceptionally robust and energy-insensitive baseline for polarization preservation. We conclude that a hybrid strategy, using a DLC snake scheme as a symmetric foundation and betatron phase tuning for fine corrections, offers the most effective path forward for the EIC and future high-energy polarized beam facilities.

\end{abstract}

\maketitle

\section{Introduction}\label{sec:1}
The scientific program of the future Electron-Ion Collider (EIC) at Brookhaven National Laboratory depends critically on the acceleration of high-intensity, high-polarization hadron beams to energies up to 275 GeV~\cite{Willeke:2021ymc, berg:ipac23-mopl156}. Preserving polarization through acceleration is a formidable challenge: as the beam is ramped, it must cross hundreds of spin-depolarizing resonances~\cite{ranjbar:pac01-rpah139}. The standard method of overcoming these resonances in high-energy synchrotrons is the use of Siberian snakes, specialized magnet insertions that rotate the spin by 180° around some axis~\cite{Derbenev:472401, Derbenev:1978hv, 10.1063/1.38281, Wienands:1990ei, syphers1997helical}. These axes are chosen to fix the stable spin precession tune at \(\nu_0\) = 1/2, which avoids all first-order spin resonances. Nonetheless, the presence of strong first-order resonances of the ring without snakes gives rise to higher-order resonant phenomena after introducing snakes. A high degree of symmetry in the accelerator lattice simplifies the resonance structure without snakes. In contrast to its predecessor, the Relativistic Heavy Ion Collider (RHIC), whose threefold super-periodicity significantly simplifies the resonance structure, the EIC Hadron Storage Ring (HSR) includes complex interaction regions (IRs) that break this symmetry~\cite{witte:ipac21-wepab002,berg:ipac23-mopl158}. The result is the excitation of significantly stronger spin resonances that can severely depolarize the beam.

To address these challenges, depolarizing spin kicks from different sections of the ring must be made to cancel, a process we refer to as ``snake matching"~\cite{hoffstaetter2004optimal,hoffstaetter2006high}. This paper systematically explores two complementary strategies for optimizing polarization transmission in the HSR. The first approach (Method I) manipulates the orbital dynamics by optimizing the vertical betatron phase advance \(\phi_y\) between snakes. By tuning the orbital phase, the depolarizing effects of different lattice sections can be made to cancel, effectively establishing a snake-matched condition across a given energy range. This approach leaves the snake configuration fixed and leverages lattice optics to suppress resonance driving terms.

The second, more powerful approach (Method II) manipulates the spin dynamics directly by varying the snake rotation axes. For a ring with \(2N\) snakes, the single constraint of \(\nu_0 = \frac{1}{2}\) leaves a \(2N-1\) dimensional parameter space of possible snake-axis configurations. Within this space, additional symmetry constraints yield systematically stronger polarization preservation. We revisit the known Lee-Courant schemes~\cite{vogt2000bounds}, which impose N constraints to achieve local spin phase advances of $\pi$, and show that these outperform generic configurations. Building on this, we introduce and analyze a novel, more restrictive arrangement we term the “Doubly Lee-Courant” (DLC) scheme. This highly symmetric configuration imposes \(2N-1\) constraints, leaving only a single degree of freedom, and enforces a \(\pi\) spin phase advance across every consecutive pair of snakes. To our knowledge, this is the first systematic investigation of the DLC class, which we find provides an exceptionally robust baseline for suppressing depolarization.

We present a comprehensive comparative analysis of these two methods for the EIC HSR. With Method I, using non-perturbative, spin-orbit simulations that compute the amplitude-dependent spin tune, we demonstrate that optimized betatron-phase schemes can successfully navigate the beam while ramping through the strongest resonant regions in realistic lattice models. We then show with Method II that snake-axis optimization, particularly under the DLC principle, provides an even more powerful route. Finally, we assess the practical trade-offs between the two approaches, considering both optics control and hardware constraints, and conclude that the symmetry principles uncovered in our snake-axis study offer a flexible and powerful new toolkit for the design of future high-energy polarized-beam facilities.

\section{Theoretical Framework and Methodology}\label{sec:2}

\subsection{Spin Dynamics in Accelerator Lattices}

\subsubsection{The Thomas-BMT Equation}
The motion of a particle's spin vector, $\mathbf{S}$, in an external electromagnetic field is described by the Thomas-Bargmann-Michel-Telegdi (T-BMT) equation~\cite{PhysRevLett.2.435}. This equation can be derived semiclassically by considering the Larmor precession in the particle's rest frame and applying the requisite Lorentz boosts to the laboratory frame, which correctly incorporates the relativistic effect of Thomas precession. Alternatively, it can be obtained as a semiclassical expansion in \(\hbar\) of the Dirac equation through a Foldy-Wouthuysen transformation (unlike the usual non-relativistic expansion seen in textbooks)~\cite{Conte:1996ad, Heinemann:1998sk}. In the lab frame, for a particle with charge $q$, mass $m$, and momentum $\mathbf{p} = \gamma m \mathbf{v}$, the spin precesses according to $\frac{d\mathbf{S}}{dt} = \boldsymbol{\Omega}_\mathrm{BMT} \times \mathbf{S}$, where the precession vector is given by:
\begin{multline}
    \boldsymbol{\Omega}_\mathrm{BMT} = -\frac{q}{\gamma m}\left[(G\gamma+1)\mathbf{B}_\perp + (G+1)\mathbf{B}_\parallel \right.\\
    \left. - \frac{G\gamma^2}{\gamma+1}\frac{\mathbf{v}}{c^2}(\mathbf{v}\cdot\mathbf{B})
    - \left(G\gamma+\frac{\gamma}{\gamma+1}\right)\frac{\mathbf{v}}{c^2}\times\mathbf{E}\right],
\end{multline}
where $G = (g-2)/2$ is the gyromagnetic anomaly, and $\mathbf{B}_\perp$ and $\mathbf{B}_\parallel$ are the magnetic field components perpendicular and parallel to the particle's velocity, respectively.

In a modern high-energy synchrotron like the EIC Hadron Storage Ring (HSR), the particle motion is ultra-relativistic ($\gamma \gg 1$) and the lattice is composed primarily of transverse magnetic fields for bending and focusing. The electric fields are localized to RF cavities and are longitudinal, and solenoidal fields are generally not used for high-energy hadron beams. Under these conditions and the paraxial approximation, the T-BMT equation simplifies considerably. Transforming to a co-moving coordinate frame that follows the particle's trajectory $s(t)$, the equation of motion becomes $\frac{d\mathbf{S}}{ds} = \boldsymbol{\Omega}(s) \times \mathbf{S}$, with the precession vector dominated by the magnetic fields perpendicular to the design orbit:
\begin{equation}
    \boldsymbol{\Omega}(s) \approx -\frac{q}{\gamma m\mathbf{v}}\left[(G\gamma+1)\mathbf{B}_\perp\right] = -\frac{q}{p}\left[(G\gamma+1)\mathbf{B}_\perp\right].
\end{equation}
The term $G\gamma$ represents the number of additional spin precessions relative to the momentum vector per turn. On the design closed orbit of a perfectly flat ring with only vertical dipole fields $B_y$, the precession is purely vertical. The rate of this precession relative to a revolution defines the \textit{closed-orbit spin tune}, $\nu_0$:
\begin{equation}
    \nu_0 = \frac{1}{2\pi} \oint \frac{qG\gamma B_y}{p} ds = G\gamma.
\end{equation}
Associated with this tune is a periodic, stable spin direction on the closed orbit, known as the \textit{invariant spin field} on the closed orbit, $\mathbf{n}_0(s)$. In an ideal flat ring, $\mathbf{n}_0$ is purely vertical, $\mathbf{n}_0(s) = \hat{\mathbf{y}}$.

\subsubsection{Spin-Orbit Resonances and Resonance Strength}
Depolarization occurs when perturbing horizontal magnetic fields cause the spin to precess away from the stable $\mathbf{n}_0$ direction. These kicks can add up coherently if the frequency of the spin precession synchronizes with the frequency at which the particle encounters these perturbations. This is the condition for a spin-orbit resonance. To analyze this, we leverage a canonical spin-orbit formalism in which we define a rotating spin frame $(\hat{\mathbf{l}}, \hat{\mathbf{m}}, \hat{\mathbf{n}}_0)$ whose vertical axis is $\hat{\mathbf{n}}_0$, the invariant spin field on the closed orbit~\cite{barber1994canonical}. $\hat{\mathbf{l}}$ and $\hat{\mathbf{m}}$ are then orthogonal vectors in the plane perpendicular to $\hat{\mathbf{n}}_0$ and precess around $\hat{\mathbf{n}}_0$ with the closed-orbit spin tune $\nu_0$. Spin kicks are driven by the component of $\boldsymbol{\Omega}$ in the $(\hat{\mathbf{l}}, \hat{\mathbf{m}})$ plane. 

For particles executing betatron oscillations around a flat closed-orbit, the primary source of these kicks are the quadrupole magnets, which have horizontal fields $B_x$ off-axis. A particle with vertical position $y(s)$ in a quadrupole with gradient $k(s) = \frac{q}{p}\frac{\partial B_x}{\partial y}$ experiences a precession kick $\propto G\gamma \cdot k(s)y(s)$. Resonances occur when the spin tune matches a harmonic of the orbital motion:
\begin{equation}
    \nu_0 = k_0 + k_x Q_x + k_y Q_y + k_s Q_s, \quad k_i \in \mathbb{Z},
\end{equation}
where $Q_{x,y,s}$ are the betatron and synchrotron tunes. The most dangerous, first-order resonances for a flat ring are the \textit{intrinsic resonances}, driven by vertical betatron motion ($|k_y|=1$):
\begin{equation}
    \nu_0 = k \pm Q_y.
\end{equation}
The strength of these resonances, $\epsilon_k$, is defined by the Fourier harmonic of the spin-perturbing fields as seen by the particle. For an intrinsic resonance, this strength is given by the spin-orbit coupling integral~\cite{hoffstaetter2006high} (also referred to as the resonance driving term):
\begin{equation}\label{q}
    \epsilon_{y} = \frac{G\gamma+1}{2\pi} \oint \sqrt{\frac{J_y \beta_y(s)}{2}} k(s) e^{i(-\psi_0(s) \mp \phi_y(s))} ds,
\end{equation}
where $J_y$ is the vertical action, $\beta_y(s)$ is the vertical beta function, $\phi_y(s)$ is the vertical betatron phase advance, and $\psi_0(s)$ is the accumulated spin precession phase on the closed orbit. For a ring with only vertical fields on the closed orbit, \(\psi_0(s) = \int_0^s \frac{G\gamma}{\rho(s')} ds'\). Non-zero resonance strength leads to rapid depolarization as the beam is accelerated through the resonance energy.

\subsubsection{The Invariant Spin Field and Amplitude-Dependent Spin Tune}
The concept of a stable spin direction can be generalized from the closed orbit to any stable trajectory in 6D phase space~\cite{derbenev1973polarization, PhysRevSTAB.7.124002}. For any particle executing stable oscillations on an orbital torus defined by actions $\mathbf{J} = (J_x, J_y, J_s)$, there exists an \textit{Invariant Spin Field} (ISF), $\mathbf{n}(\mathbf{J}, \boldsymbol{\phi})$, which is periodic in the orbital angle coordinates $\boldsymbol{\phi}$. A spin aligned with $\mathbf{n}$ at a point in phase space will remain aligned with $\mathbf{n}$ as the particle moves. The deviation of $\mathbf{n}(\mathbf{J}, \boldsymbol{\phi})$ from the closed-orbit direction $\mathbf{n}_0$ determines the maximum achievable polarization for an ensemble of particles with actions $\mathbf{J}$. This is given by the phase average of the ISF:
\begin{equation}
    P_\text{lim}(\mathbf{J}) = |\langle \mathbf{n}(\mathbf{J}, \boldsymbol{\phi}) \rangle_{\boldsymbol{\phi}}|.
\end{equation}
Associated with the ISF is the \textit{Amplitude-Dependent Spin Tune} (ADST), $\nu(\mathbf{J})$, which is the effective spin precession tune on that particle's trajectory~\cite{mane2002exact}. Near the closed orbit, the ADST can be Taylor-expanded as a function of the actions~\cite{forest2016tracking}:
\begin{equation}
    \nu(\mathbf{J}) = \nu_0 + \nu_{1x} J_x + \nu_{1y} J_y + \dotsb
\end{equation}
The linear coefficients, particularly $\nu_{1y}$ for a flat ring, quantify the strength of the resonance driving terms across the whole machine. A large value of $\nu_{1y}$ indicates that particles with larger betatron amplitudes have spin tunes that deviate significantly from the central tune, leading to a tune spread that can straddle resonance conditions and cause depolarization. Consequently, minimizing $|\nu_{1y}|$ is a primary and computationally efficient target for snake matching optimization.

\subsection{Siberian Snakes and Polarization Transmission}
To overcome the myriad of first-order spin resonances crossed during acceleration, high-energy synchrotrons employ \textit{Siberian snakes}. A snake is a sequence of magnets (typically helical dipoles) that rotates the spin by $\pi$ radians around a horizontal axis, while leaving the particle orbit outside the insertion minimally perturbed~\cite{Derbenev:472401, Derbenev:1978hv, 10.1063/1.38281, Wienands:1990ei, syphers1997helical}.

The spin transport through any section of the accelerator can be represented by a $2 \times 2$ SU(2) matrix. A rotation by angle $\theta$ around an axis defined by the unit vector $\hat{\mathbf{a}} = (a_x, a_y, a_z)$ is given by $\mathcal{R}(\theta, \hat{\mathbf{a}}) = \exp(-i\frac{\theta}{2} \boldsymbol{\sigma} \cdot \hat{\mathbf{a}})$, where $\boldsymbol{\sigma}$ are the Pauli matrices. A snake with a rotation axis in the horizontal plane making an angle $\varphi$ with the longitudinal direction \(\hat{\mathbf{z}}\) has the SU(2) representation:
\begin{equation}
    \mathcal{S}(\varphi) = -i(\sin\varphi\sigma_x+\cos\varphi\sigma_z).
\end{equation}
By inserting an even number, $2N$, of snakes into the ring, the closed-orbit spin tune $\nu_0$ can be fixed at $1/2$, independent of the beam energy $G\gamma$. If the total spin precession in the arcs between snakes is canceled, the one-turn map on the closed orbit is a product of snake matrices. The condition $\nu_0=1/2$ imposes a single constraint on the $2N$ snake axis angles $\varphi_i$:
\begin{equation}
    \sum_{i=1}^{2N} (-1)^{i-1} \varphi_i = \frac{\pi}{2} \pmod{\pi}.
\end{equation}
This leaves $2N-1$ degrees of freedom in the choice of snake axes. This freedom is the basis of snake-axis optimization, a subset of ``snake matching", for which the goal is to cancel the net effect of spin-perturbing fields.

\subsection{Resonance Driving, Symmetry Breaking, and the EIC Challenge}
The efficacy of any snake matching scheme is determined by the structure of the accelerator lattice. In a ring with a high degree of super-periodicity $P$, the resonance driving terms from each super-period interfere. This leads to a systematic cancellation of all resonances except those whose integer component is a multiple of $P$. That is, only the intrinsic resonances $\nu_0 = mP +\mathbf{m}\cdot\mathbf{Q}$, $\mathbf{m}\in\{-1,0,+1\}^3$ are strongly driven.

RHIC, the EIC's predecessor, has an approximate 3-fold super-periodicity. This symmetry plays a crucial role in passively suppressing 2/3 of spin resonances, simplifying the task of polarization preservation. The EIC HSR, in contrast, inherits the RHIC tunnel but incorporates at least one new, large, and complex IR that fundamentally breaks this 3-fold symmetry.

\begin{figure*}[htbp]
    \centering
    \includegraphics[width=\textwidth]{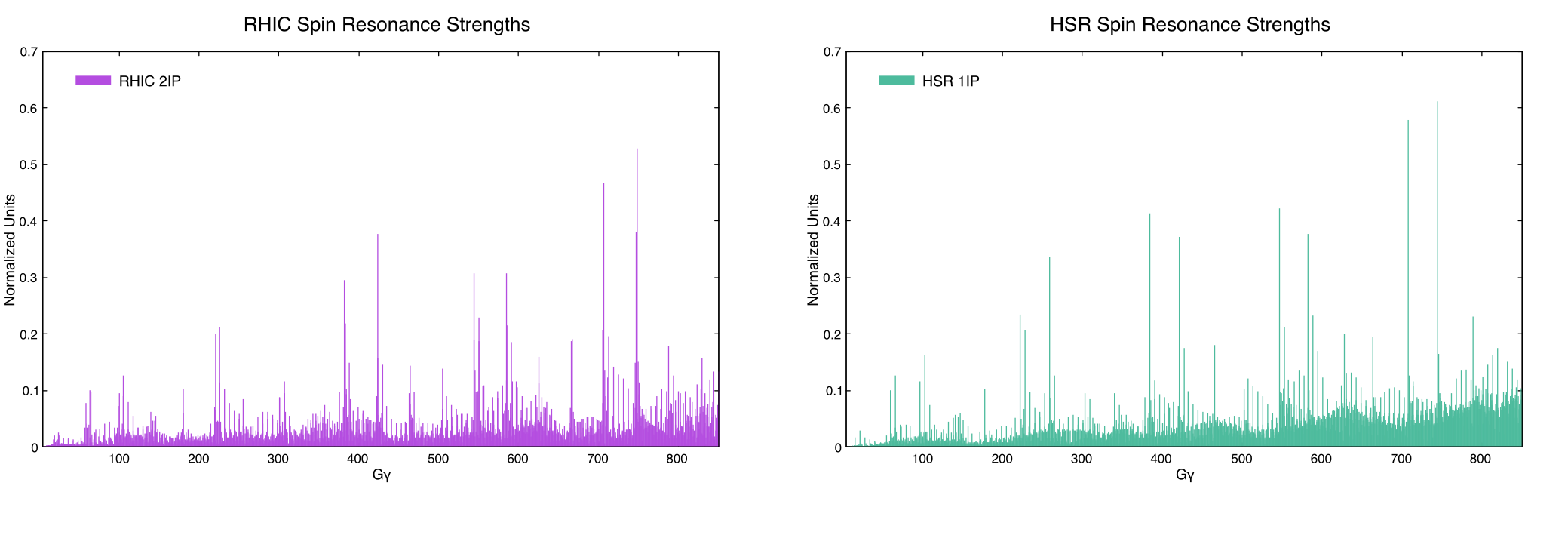}
    \caption{This schematic compares the intrinsic resonance spectrum of the more symmetric lattice of RHIC with the denser, less symmetric lattice of the HSR, which exhibits both stronger dominant resonances due to its lower $\beta^*$ and a higher resonance floor due to its symmetry-breaking IR. Despite RHIC having two low-beta interaction points compared to the single one in the HSR, its overall lattice symmetry is preserved to a much greater degree.}
    \label{fig:res_spec}
\end{figure*}

To understand this mathematically, the total resonance strength $\epsilon$ can be expressed as a sum of contributions from all insertions (arcs, IRs, etc.) around the ring, similar to Equation~\eqref{q}:
\begin{equation}
    \epsilon \propto \sum_{j=1}^{M} \int_{\text{insertion } j} f(s) e^{i\chi_j(s)} ds = \sum_{j=1}^{M} \mathcal{F}_j e^{i\bar{\chi}_j},
\end{equation}
where $f(s)$ contains the lattice functions ($\beta_y, k$) and $\chi_j(s)$ is the total spin-orbital phase advance. In a symmetric lattice, the complex $\mathcal{F}_j$ are identical for similar sections (e.g., all arcs). They get multiplied by different phases $\bar{\chi}_j$ because each section has a different initial spin-orbit phase. The sum over these phase terms can lead to constructive or destructive interference. Some resonances that are relatively much stronger than others, called systematic resonances, are formed by constructive interference from all FODO cells in the arcs. Systematic resonances produce the extremely sharp, local peaks in Figure~\ref{fig:res_spec}. On the other hand, some of the destructive interference or cancellations are lost because the EIC HSR's IR introduces very large and distinct $\mathcal{F}_j$ terms that are not balanced by other insertions. This broken symmetry means that a dense spectrum of strong, first-order, intrinsic resonances is excited, as shown in the right plot of Figure~\ref{fig:res_spec}. This makes polarization preservation in the HSR a qualitatively more difficult problem than in RHIC, demanding a more sophisticated and global approach to snake matching~\cite{hock:ipac24-tups05, hamwi:ipac24-tups33, hamwi:napac2025-tup077}. An alternative to minimizing these resonance driving terms directly is instead minimizing the deviation of the invariant spin field from the vertical axis, or more directly, minimizing the amplitude-dependent spin tune spread coefficient, $\nu_{1y}$. The latter approach prevents the excitation of higher-order resonances by avoiding higher-order resonance conditions. We follow both approaches in Section~\ref{sec:4}

\subsection{Simulation and Optimization Tools}
To analyze and compare snake matching strategies, we rely on numerical tools. The accelerator modeling library \texttt{Bmad} is used for lattice definition and particle tracking~\cite{Sagan:Bmad2006}. Non-perturbative spin dynamics are studied by tracking a distribution of particles and their spins through a detailed model of the HSR lattice.

The primary figure of merit for polarization preservation is the polarization transmission, $P_\text{dyn}$, calculated by averaging the final polarization of an ensemble of particles with action equal to a realistic beam emittance and initialized on its ISF after tracking for many turns while ramping at a realistic rate, then dividing by the final $P_\text{lim}$. 
\begin{equation}
    P_\text{dyn}\,P_\text{lim} \equiv \left|\langle \mathbf{S}\rangle_i\right|
\end{equation}
However, direct optimization of $P_\text{dyn}$ is computationally intensive. A cheap but indirect approach is to calculate $\epsilon_y$ at various resonances, which strongly contribute to $P_\text{lim}$. While this does not directly address $P_\text{dyn}$, it does inform the severity of first-order resonant depolarization at each energy. A more direct, but expensive proxy for optimization is the amplitude-dependent spin tune spread. Techniques such as stroboscopic averaging or normal form analysis via \texttt{Bmad/PTC} are employed to compute the coefficient $\nu_{1y}$ directly from the one-turn map of the lattice~\cite{sagan:ipac18-mopmf028, Yokoya:1999ip, forest2016tracking}. The optimization goal is then to find lattice parameters (quadrupole strengths or snake axes) that minimize $\epsilon_y$ and $|\nu_{1y}|$ over the required energy range, thereby ensuring the spin tune is stable across the entire beam phase space and suppressing the underlying drivers of depolarization. 

A modern upgrade of Bmad, SciBmad~\cite{scibmad,signorelli:napac2025-mop044}, has been written in the high-performance, scientific programming language Julia. SciBmad is already well-developed and offers many tools. These include complex beamline construction, GPU-parallelized high-order symplectic integration of static electromagnetic fields, high-order truncated power series computation via automatic differentiation, and nonlinear normal form map analysis tools for both orbital and spin-orbital maps. Altogether, these packages were confirmed to replicate many of the results of this study to within machine precision.

\section{The Polarization Challenge in the EIC Hadron Storage Ring}\label{sec:3}

The scientific mission of the EIC necessitates a hadron storage ring that has the capability of accelerating polarized beams of various species to high energies, including protons (p), deuterons (D), helions $\left(^3He^{2+}\right)$, and possibly lithium-6 $\left(^6Li^{3+}\right)$ and lithium-7 $\left(^7Li^{3+}\right)$ nuclei~\cite{ZELENSKI2023168494, PhysRevAccelBeams.23.021001, peng2024}. While the accessible energy range for these beams is constrained by the maximum fields of the superconducting accelerator $B_\text{max}\rho\geq\frac{p}{q}$, their anomalous magnetic moments $G\equiv(g-2)/2$ are responsible for the extent of depolarization they must withstand. In the absence of Siberian snake devices, ramping the beam energy through each integer of $G\gamma$ subjects the beam to one imperfection resonance plus two intrinsic resonance sidebands, shown in Table~\ref{tab:species}.

\begin{table}[H]
\centering
\caption{Species-dependent Linear Resonance Spectrum}
\label{tab:species}
\begin{tabular}{|c|c|c|c|}
\hline
\textbf{Species} & \textbf{G} & \textbf{max $|G\gamma|$} & \textbf{No. of Resonances}\\ 
\hline
$p$ & $\quad1.7928\,$ & 525 & 1575 \\
$D$ & $\,-0.1430\,$ & 21 & 63 \\
$^3 He^{2+}$ & $\,-4.1842\,$ & 819 & 2457 \\ 
$^6 Li^{3+}$ & $\,-0.1818\,$ & 27 & 81 \\ 
$^7 Li^{3+}$ & $\quad1.5196\,$ & 191 & 573 \\ 
\hline
\end{tabular}
\end{table}

As seen in Equation~\eqref{q}, the spin resonance strength roughly scales with $\sqrt{G\gamma}$ when taking into account the adiabatic damping of beam size with energy. Thus, we expect that as the number of resonances in a ramp increases (with energy), so does the strongest possible resonance. Therefore, helion beams are subjected to the worst possible depolarizing conditions, and so we only consider those beams in this study. On the other hand, particles with small anomalous magnetic moment $\left(^2 H,\;^6 Li\right)$ are not amenable to Siberian snake devices, since their small $G$-factors require impractically long snakes. Similarly to protons in the AGS, the polarization of these beams will have to be handled with partial Siberian snakes. These beams will not be discussed further in this study.

For our baseline studies, we adopt the nominal HSR lattice configuration with six Siberian snakes placed in the straight sections between the six arcs. This arrangement, with two snakes per ``super-period" of the old RHIC layout and $\pi/3$ radians of revolution between each snake, ensures an energy-independent closed-orbit spin tune of $\nu_0 = 1/2$. 

In the first study of betatron phase matching, snake axes are configured in an alternating pattern of $\varphi_i = \pm 15^\circ$ with respect to the longitudinal beam direction. This choice offers the advantage of all snakes being identical up to handedness, as well as offering minimal orbit excursion within and outside the snakes. In Section~\ref{sec:4}, we show that this choice is completely inadequate and that further modifications must be made. We find in Section~\ref{sec:5} that the configuration of \(\varphi_i = \pm45^\circ\), a natural extension of the RHIC operating snakes, is a more promising alternative, albeit with larger orbit excursions. Furthermore, we introduce a new parameter to find the optimal snake configuration.

\begin{figure}[h]
    \centering
    \includegraphics[width=\linewidth]{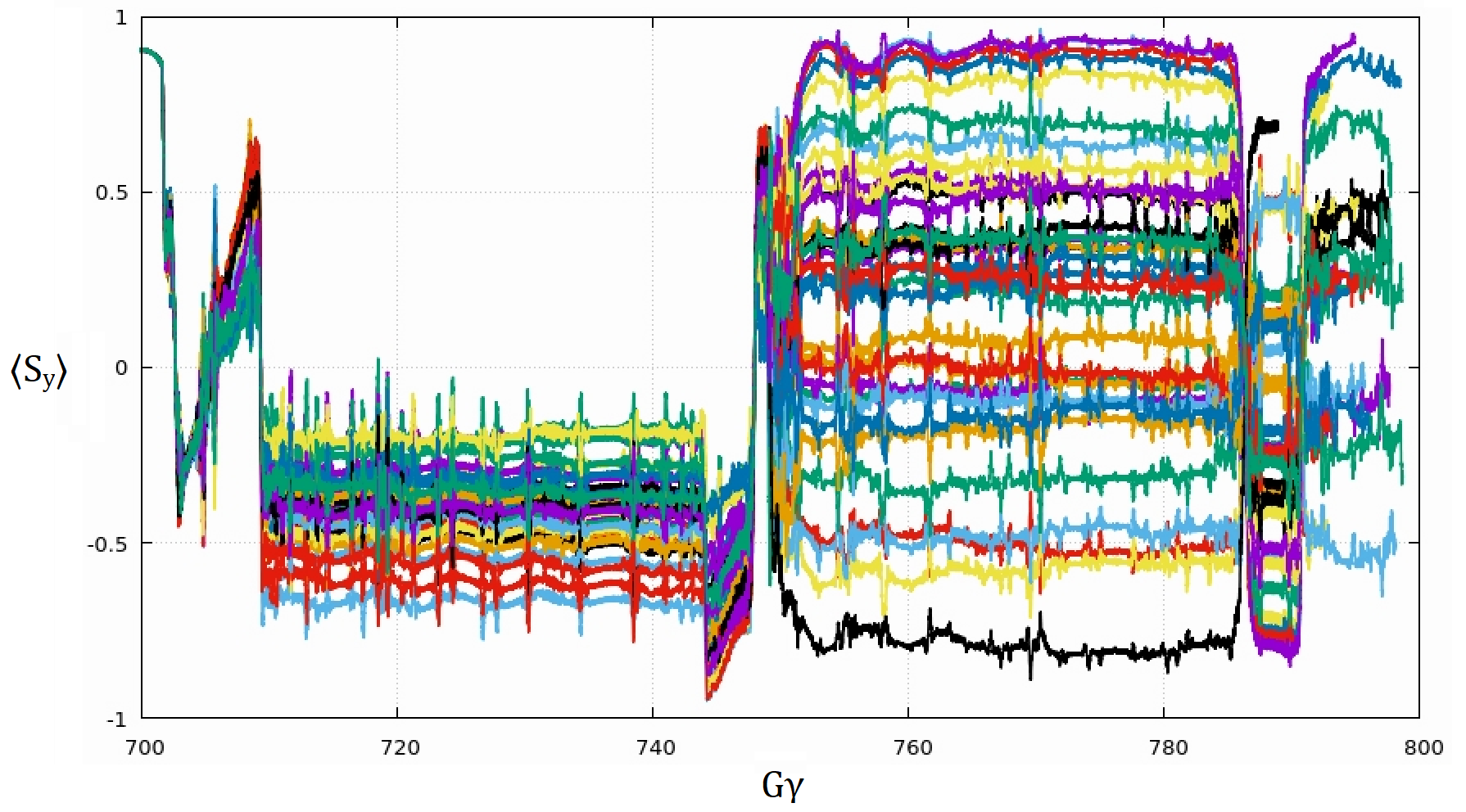}
    \caption{Baseline spin tracking of uncooled, round, helion beams of \(3\,\mu\text{m}\) emittance in the injection HSR lattice used for ramping with $\pm15^\circ$ snakes. Each color represents the vertical spin component of each particle in the bunch. The ensemble polarization is lost rapidly as the beam is accelerated through strongly resonant energies.}
    \label{fig:track_og}
\end{figure}

We demonstrate the inadequacy of polarization transmission for the $\pm15^\circ$ solution via non-perturbative spin tracking simulations. A beam of 49 helium-3 nuclei (helions) is initialized on a 4D phase-space torus with amplitudes corresponding to the beam periphery ($J_x = \sigma_x$, $J_y = 2\sigma_y$) to probe the depolarizing effects on large vertical-amplitude particles, which are typically the first to depolarize in a flat ring. The initial spin of each particle is aligned with its local ISF. The beam is then accelerated at a realistic rate of \(\dot \gamma=0.75/\text{s}\) through the most problematic energy range ($700 \le G\gamma \le 800$), where significant polarization loss is observed. Figure \ref{fig:track_og} shows the result of this baseline tracking. The polarization rapidly deteriorates as the beam crosses strongly resonant regions, particularly around $G\gamma \approx 705$ and $G\gamma \approx 745$, confirming that this snake setup is insufficient to preserve polarization in the face of the HSR's strong non-systematic resonances.

\section{Method I: Optimization of Betatron Phase Advance}\label{sec:4}

The first strategy we investigate for restoring polarization transmission is the manipulation of the accelerator's orbital dynamics, specifically the vertical betatron phase advance $\Phi_y$ between the Siberian snakes. This approach treats the snake configuration as fixed (e.g., the $\pm15^\circ$ axes) and instead adjusts the lattice optics to force the depolarizing spin kicks from different sections of the ring to cancel one another.

\subsection{Theoretical Basis for Phase Advance Optimization}
The core idea is to treat the total first-order resonance strength, represented by the complex-valued spin-orbit coupling integral $I^\pm$ in Eq.~\eqref{q}, as a vector sum in the complex plane. The ring is composed of $2N$ sections between the snakes, and the total integral is the sum of the contributions from each section:
\begin{equation}
    I^\pm = \sum_{j=1}^{2N} I_j^\pm.
\end{equation}
The integral for the $j$-th section, $I_j^\pm$, accumulates a certain magnitude and phase determined by its specific lattice functions and length. A Siberian snake fundamentally alters the spin precession that follows it. As derived in Section~\ref{sec:2}, passing through the $i$-th snake transforms the spin basis $(\mathbf{m}+i\mathbf{l})$ to $e^{i2\varphi_i}(\mathbf{m}-i\mathbf{l})$ and inverts the subsequent vertical spin precession. Consequently, the phase of the complex number $I_j^\pm$ depends on the accumulated spin phase $\Psi$ and betatron phase $\Phi_y$ up to that point, including the discrete jumps from the snakes. Schematically, the contribution from section $j$ can be written as:
\begin{equation}
    I_j^\pm \propto e^{i\chi_j^\pm} \int_{\text{section } j} k(s)\sqrt{\beta_y(s)} e^{i(-\psi_j(s) \pm \phi_j(s))} ds,
\end{equation}
where $\psi_j(s)$ and $\phi_j(s)$ are the phases within section $j$, and the global phase factor $\chi_j$ is a function of the preceding snake axes and the total inter-snake phase advances:
\begin{equation}
    \chi_j^\pm = \sum_{i=1}^{j-1}\left[(-1)^{i-1}(\Psi_i+2\varphi_i) \pm \Phi_{y,i}\right].
\end{equation}
The snake match condition is achieved when the sum of these complex vectors is zero:
\begin{equation}
    \sum_{j=1}^{2N} I_j^\pm = 0.
\end{equation}
Since the snake axes $\varphi_i$ and arc precession angles $\Psi_i$ are fixed in this method, the only free parameters are the betatron phase advances $\Phi_{y,i}$ between the snakes. By adjusting the quadrupole settings in the arcs and straight sections, we can change the values of $\Phi_{y,i}$, effectively "rotating" each phasor $I_j^\pm$ in the complex plane until their sum cancels. However, because the HSR lattice is not symmetric, the magnitudes $|I_j^\pm|$ are all different, a simple, symmetric choice of phase advances will not work. A numerical optimization is required to find a set of $\{\Phi_{y,i}\}$ that satisfies this condition, typically over a specific energy range where a particular resonance is strong.

\subsection{Baseline Behavior and Tune Scans}
\begin{figure}[H]
    \centering
    \includegraphics[width=\linewidth]{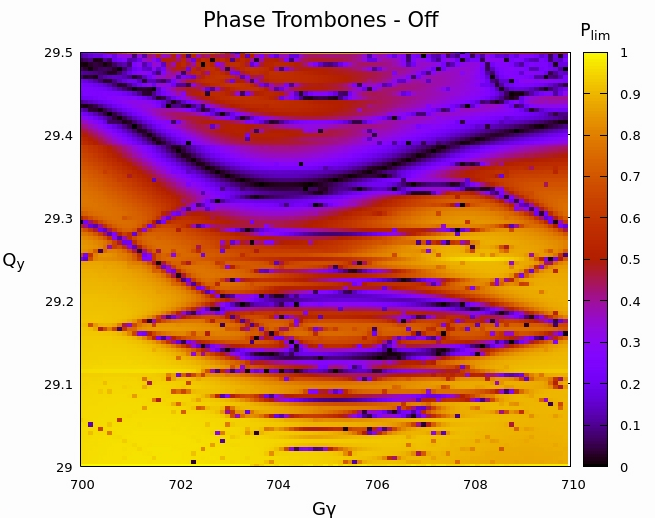}
    \caption{Tune scan of the limiting polarization $P_\text{lim}$ for the baseline HSR lattice at $G\gamma \approx 704$. The dark bands indicate strong higher-order resonance doublets. The nominal working tune is marked.}
    \label{fig:scan_og}
\end{figure}

Before optimization, we diagnose the lattice by performing a ``tune scan," where we compute the limiting polarization $P_\text{lim}$ for large-amplitude particles as a function of the vertical betatron tune $Q_y$ at a fixed energy. Figure \ref{fig:scan_og} shows such a scan for the nominal HSR lattice. The yellow areas indicate high polarization ($P_\text{lim} \approx 1$), while the dark lines indicate strong depolarization. These lines are higher-order resonance doublets, which arise from the condition $\nu(J) = k \pm m Q_y$. The splitting of the doublets is a direct measure of the deviation of the ADST from the central tune, i.e., the strength of the coefficient $\nu_{1,y}$. The wider the split, the stronger the depolarization.

\subsection{Optimization Results}
\begin{figure}[H]
    \centering
    \includegraphics[width=\linewidth]{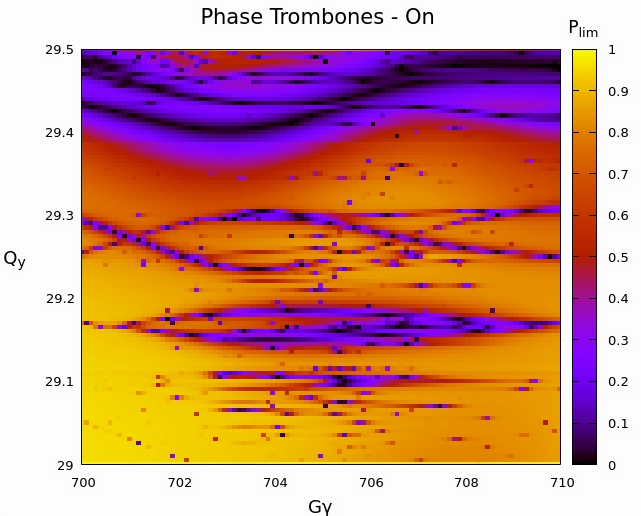}
    \caption{Tune scan for the HSR lattice after optimizing the inter-snake betatron phase advances. The resonance doublets are significantly narrowed, creating a larger area of high polarization.}
    \label{fig:scan_opt1}
\end{figure}

Our optimization procedure uses the \texttt{Bmad}-based optimizer \texttt{Tao} to vary the strengths of quadrupoles in the lattice. Initially, we model the phase adjustments with idealized ``phase trombone" \(4\times4\) matrix elements. The optimization target is to minimize $I^\pm$ as a proxy for $|\nu_{1,y}|$, which corresponds to closing the resonance doublets seen in the tune scan. Figure \ref{fig:scan_opt1} shows the tune scan after optimizing the phase advances at $G\gamma \approx 704$. The optimization has successfully squeezed the doublet lines together, creating a much larger region of stable polarization transmission.

\begin{figure}[H]
    \centering
    \includegraphics[width=\linewidth]{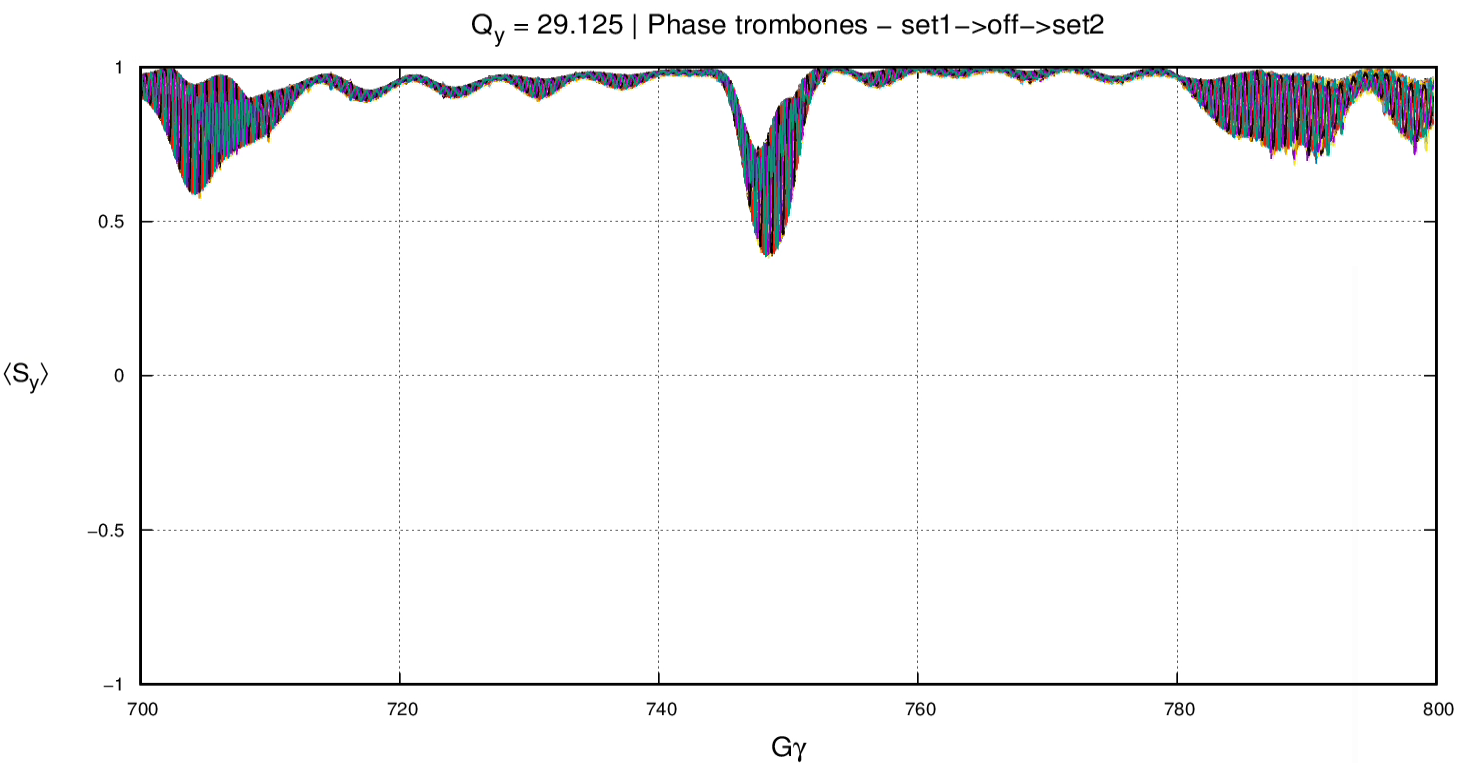}
    \caption{Spin tracking with an energy-dependent interpolation of optimized phase advances. Polarization is successfully preserved through the this set of the ramp. \(P_\text{dyn} > 99\%\)}
    \label{fig:track_old_2}
\end{figure}

This improved lattice, however, is optimized for a single energy. Since the spin-orbital phasing is energy-dependent, a solution that works at one energy may not work at another. To achieve high polarization throughout the acceleration ramp, the phase advances must be dynamically adjusted. We identify several key energy points corresponding to the strongest resonances and find an optimized set of phase advances for each. By interpolating the required quadrupole settings between these points, we can create a continuous `phase ramp." Beyond this, some tweaking of the working point is necessary for safe passage between the resonance doublet bands. Figure \ref{fig:track_old_2} shows the result of tracking through the full energy range using such an interpolated ramp. Polarization transmission is now successfully maintained with \(P_\text{dyn}\geq 99\%\).

\begin{figure}[H]
    \centering
    \includegraphics[width=\linewidth]{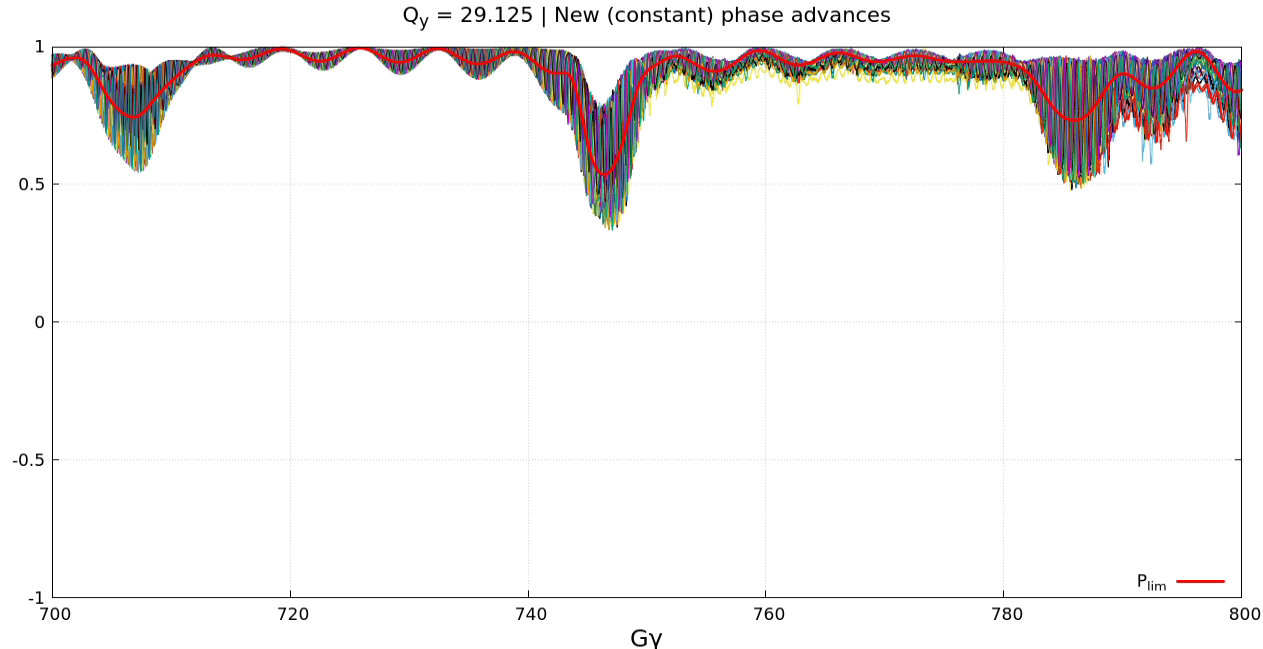}
    \caption{Spin tracking using a single, constant set of optimized phase advances, found via multi-objective optimization. This provides a robust solution without requiring dynamic lattice changes during the ramp. \(P_\text{dyn}> 96\%\)}
    \label{fig:plot7}
\end{figure}

A dynamic ramp of many quadrupole circuits can be challenging to implement, and this issue will be addressed in the following subsection. A more practical solution would be to find a single, fixed set of phase advances that provides adequate performance over the entire energy range. Using scalarized multi-objective optimization, simultaneously targeting the minimization of $I^\pm$ at several key energies, we had trouble converging to an adequate solution. Since $I^\pm$ was being used as a proxy for $|\nu_{1,y}|$, we instead leveraged \texttt{Bmad/PTC} to directly minimize $|\nu_{1,y}|$ at multiple energies and found a robust, constant-phase solution. As shown in Figure \ref{fig:plot7}, this single setting also preserves polarization to a high degree, albeit with slightly larger oscillations than the fully dynamic ramp. In the end \(P_\text{dyn} > 96\%\).

\subsection{Practical Implementation and Limitations}

The idealized phase trombones must be realized by adjusting the currents in the actual HSR quadrupoles. This involves a complex lattice matching procedure, as changing quadrupole strengths affects not only the phase advance but also the beta functions and dispersion, which must be matched at centers of the straight sections. Furthermore, adjusting quadrupoles to control $\Phi_y$ inevitably causes a correlated change in the horizontal phase advance $\Phi_x$, which is undesirable and could hypothetically excite new coupling resonances. To achieve this control, the phase advance in each arc is made adjustable as a function of an additional current parameter applied to the arc quadrupoles. Each arc is independently matched to its IRs by solving for unique and smooth optics configurations that satisfy $(\beta_x,\beta_y,\alpha_x,\alpha_y,\eta_x,\eta_x')$ at both IPs. The optics match is performed sequentially for a range of magnet strengths $\pm10\%$ around their nominal value, assuming a linear relation between current and magnetic field (unsaturated magnets). Different sets of matching quadrupoles are used for positive and negative changes in arc phase advance, and the resulting matched quadrupole currents are saved and cubically interpolated to provide a piecewise smooth modulation of the phase advances in each arc.

\begin{figure}[H]
    \centering
    \includegraphics[width=\linewidth]{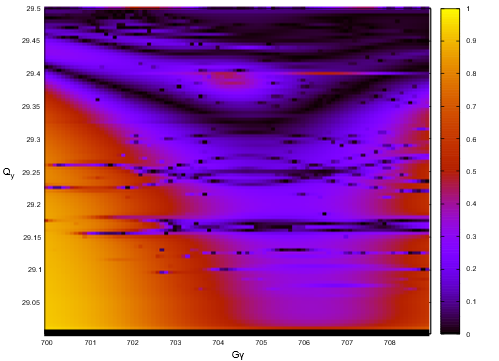}
    \caption{Tune scan after optimization using realistic quadrupole matching. The vertical resonance doublets are narrowed, but the tune space is complicated by horizontal-vertical coupling.}
    \label{fig:optimized_scan}
\end{figure}

\begin{figure}[H]
    \centering
    \includegraphics[width=\linewidth]{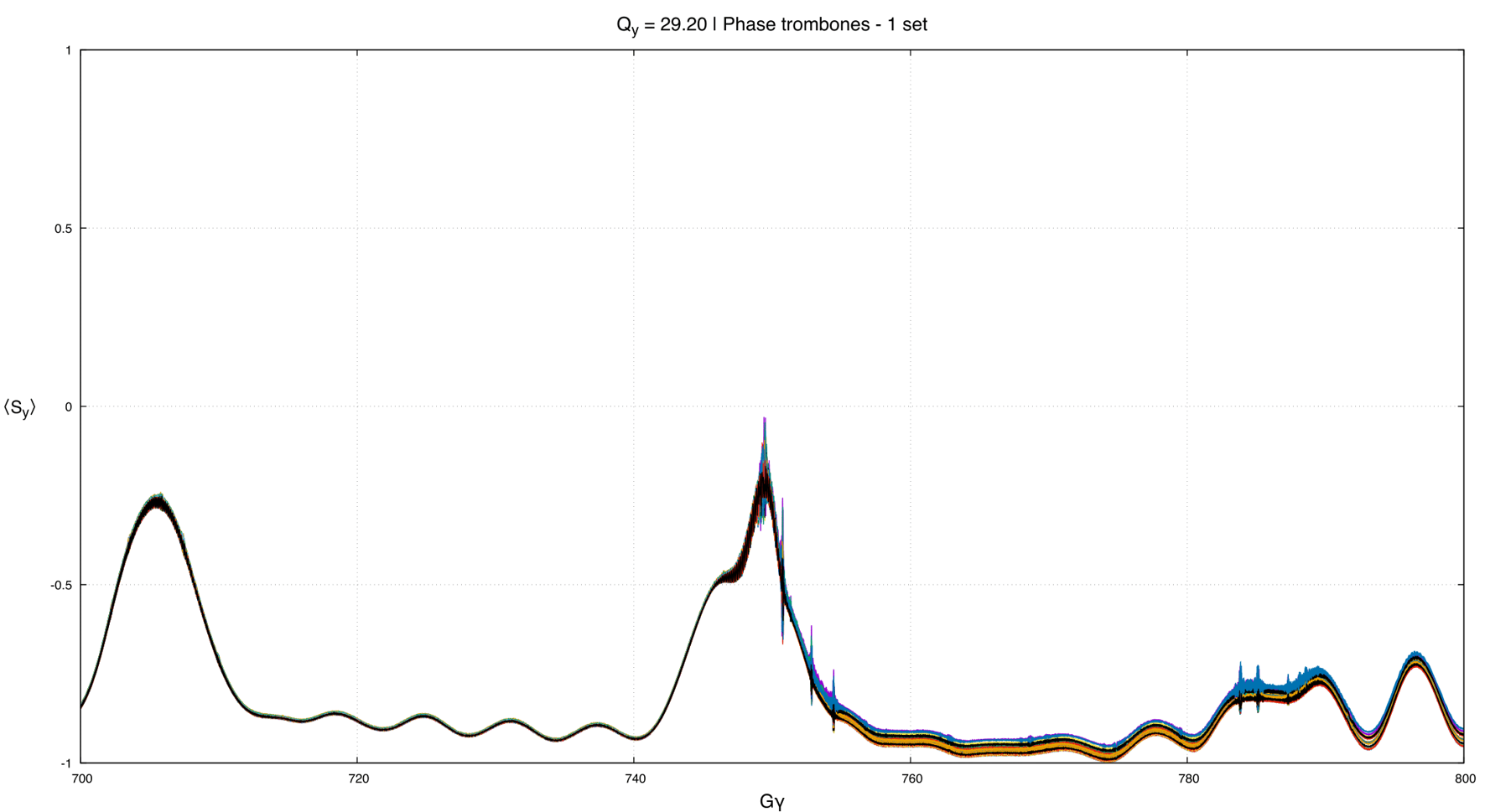}
    \caption{Final tracking result with a single, optimized set of quadrupole currents in the fully matched lattice with \(P_\text{dyn}>97\%\).}
    \label{fig:track_new_3}
\end{figure}

After developing a realistic matching procedure using the IR quadrupoles, we re-optimized the lattice by targeting the minimization of $|\nu_{1,y}|$. The resulting tune scan (Figure \ref{fig:optimized_scan}) shows that while the target vertical resonance doublets are narrowed at one strongly resonant region, the tune space shows lower average \(P_\text{lim}\) due to further breaking of the approximate 3-fold symmetry of the lattice. Nonetheless, a new safe working point can be identified. Final tracking simulations with this fully-matched, realistic lattice and a single, constant set of optimized quadrupole settings confirm the viability of the approach, preserving $\sim97\%$ polarization through the most difficult part of the energy ramp (Figure \ref{fig:track_new_3}).

The primary limitation of the betatron phase optimization method is its nature as a numerical, energy-dependent compensation. Especially for a ring primarily composed of superconducting magnets, rapid, time-dependent changes to the currents flowing through the superconducting coils drastically increases the likelihood of quenches. To avoid this, very stringent requirements on the acceleration of the ramping rate must be followed. Furthermore, it does not eliminate the source of the depolarizing kicks but rather carefully arranges for them to cancel their contribution to ADST spread at a specific energy or over a specific range. While effective, the solution can be sensitive to lattice errors and requires either complex, dynamic control of quadrupole currents or a carefully crafted compromise solution that may not be optimal at all energies. This motivates the search for a more powerful solution that suppresses the resonance driving terms at their source.

\section{Method II: Optimization of Snake Rotation Axes}\label{sec:5}

The second, more fundamental strategy for achieving polarization transmission involves manipulating the spin dynamics directly by optimizing the rotation axes of the Siberian snakes. Instead of compensating for depolarizing kicks using the lattice optics, this method aims to suppress the resonance driving terms at their source by imposing a higher degree of symmetry on the spin transport itself. We investigate this method in the 1-IP HSR lattice, where \(\beta_x^* = 80\,\text{cm},\,\beta_y^* = 7.2\,\text{cm}\) at IP6. These miniscule values of \(\beta_{x,y}^*\) give rise to \(\beta_{x,y} > 1000\,\text{m}\) in the final focusing system, which is challenging for polarization due to the larger depolarizing quadrupole kicks. Even more challenging, in comparison with RHIC, is the asymmetric interaction region, where \(\beta_{y,\text{max}} \approx 550\,\text{m}\) upstream of the IP and \(\beta_{y,\text{max}} \approx 1200\,\text{m}\) downstream of the IP. This breaking of the approximate symmetry further enhances the non-systematic resonances as described previously.

\subsection{Exploration of the General Snake Axis Space}
For a ring with $2N$ snakes, the condition to fix the closed-orbit spin tune at $\nu_0 = 1/2$ imposes only a single constraint on the $2N$ snake rotation axes $\{\varphi_i\}$. This leaves a vast $2N-1$ dimensional parameter space of possible configurations. A practical consideration often limits this space: snakes with axes far from the longitudinal direction (i.e., closer to radial, $\varphi = \pm 90^\circ$) generate larger orbit excursions, which can reduce the dynamic aperture and enhance beam instabilities. For this reason, configurations with minimal deviation from the longitudinal axis, such as the $\pm15^\circ$ scheme for the HSR's six snakes, can be preferred. Details about the orbital and optical properties of snake axes for different beams are explored in~\cite{snakes}.

However, an orbitally-optimal solution is not necessarily optimal for polarization. We perform a statistical survey over the 5-dimensional snake axis parameter space for the HSR over 49 equidistant beam energies in the \(G\gamma \in (700,800)\) range. We consider the maximum ADST deviation \(\max_{G\gamma}(\Delta\nu)\) to non-perturbatively quantify the doublet splitting as well as \(\min_{G\gamma}(P_\text{lim})\) to quantify the mean value of the residual resonance driving terms.

Sampling almost 250,000 unique snake configurations reveals that the $\pm15^\circ$ scheme is extremely inadequate: it is far from the best-performing configuration in terms of \(\max(\Delta\nu)\), \(\min(P_\text{lim})\), and also \(P_\text{dyn}\) when tracking a bunch. Many other solutions, albeit with potentially larger orbit excursions, offer significantly improved spin stability. This motivates a search for underlying principles that identify these superior configurations.

\subsection{The Statistical Advantage of Lee-Courant Schemes}
A known class of symmetric configurations, which we refer to as Lee-Courant (L-C) schemes, imposes additional constraints on the snake axes. In a ring with $2N$ snakes, $N$ odd, an L-C scheme constrains the spin transport across pairs of snakes to be a pure $\pi$ rotation. For example, in the HSR with six snakes, this requires the one-turn map sections from snake 1 to 3, from 3 to 5, and from 5 back to 1 to each correspond to a spin tune of 1/2. This imposes $N=3$ independent constraints on the six snake axes, reducing the dimensionality of the solution space from five to three.

\begin{figure}[H]
    \centering
    \includegraphics[width=1.1\linewidth]{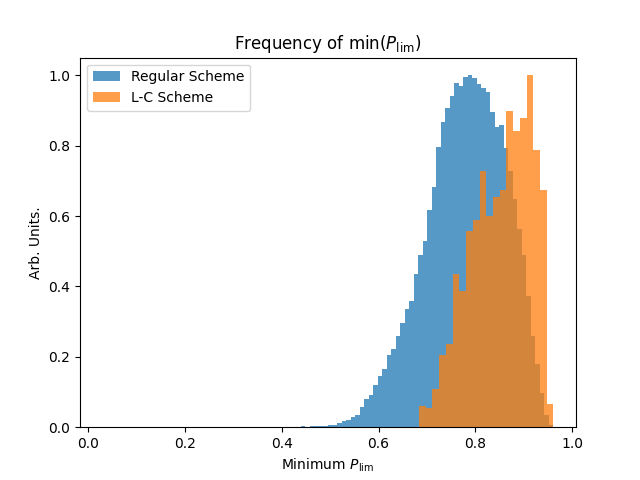}
    \caption{Histogram of \(\min(P_\text{lim})\) for a random sampling of general snake axis configurations (blue) versus the subset satisfying the Lee-Courant conditions (orange). The L-C schemes are statistically far more likely to have higher equilibrium polarization.}
    \label{fig:lc_plim_histogram}
\end{figure}

Figure~\ref{fig:lc_plim_histogram} compares the distribution of \(\min_{G\gamma}(P_\text{lim})\) from our statistical scan for the general set of solutions versus the subset that satisfies the L-C conditions. The result is unambiguous: the L-C schemes show a strong statistical preference for high polarization. This demonstrates that imposing additional symmetries on the spin transport systematically improves polarization preservation.

\subsection{The Doubly Lee-Courant (DLC) Scheme}
Building on this principle, we introduce and analyze a novel, more restrictive arrangement which we term the ``Doubly Lee-Courant" (DLC) scheme. The DLC scheme elevates the concept of local spin cancellation to its logical conclusion by enforcing a $\pi$ spin phase advance across \textit{every consecutive pair} of snakes.

\begin{figure}[H]
    \centering
    \includegraphics[width=1\linewidth]{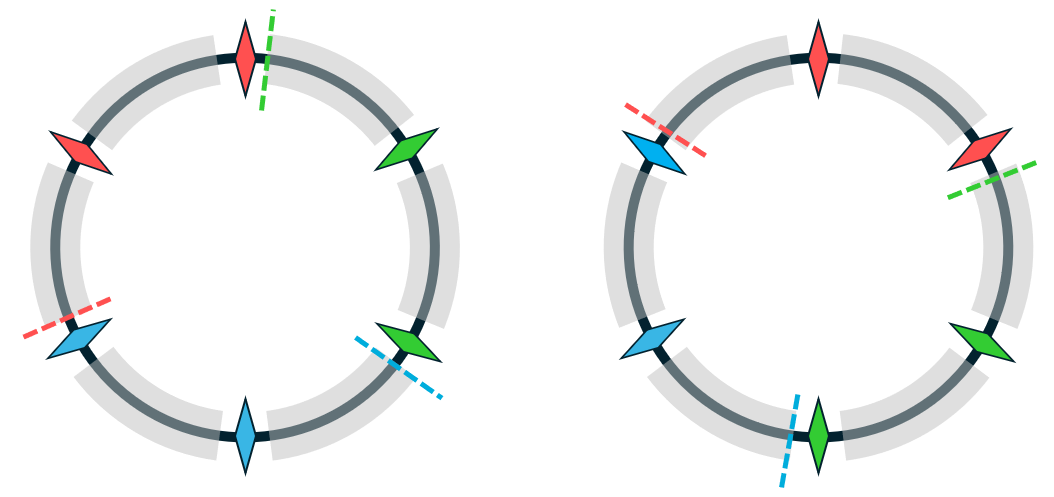}
    \caption{Schematic of a Doubly Lee-Courant snake configuration. Snakes are diamond shapes, and complete segments begin at dashed lines of one color and end at the second snake of the same color, going clockwise. Note that each snake simultaneously participates in two color schemes, i.e. cancellation in two overlapping segments.}
    \label{fig:dlc_schematic}
\end{figure}

To this end, each snake simultaneously participates in fixing the spin phase advance across two overlapping segments which begin at its upstream \emph{and} end at its downstream neighbor, as shown in Fig.~\ref{fig:dlc_schematic}. This imposes the highest possible degree of local symmetry on the spin motion.

\begin{figure}[H]
    \centering
    \includegraphics[width=1.1\linewidth]{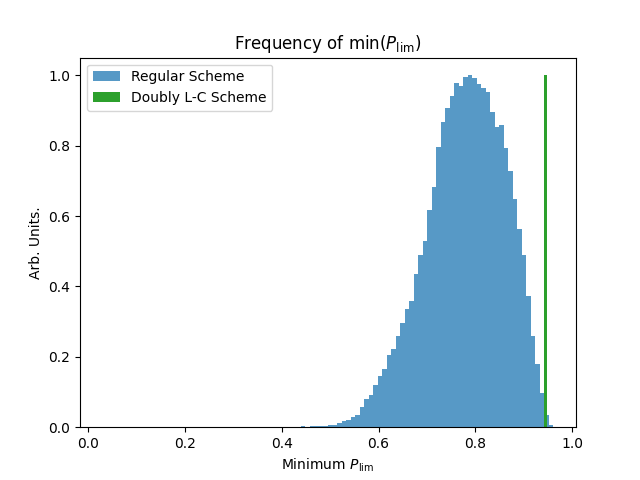}
    \caption{Histogram of \(\min(P_\text{lim})\) for a random sampling of general snake axis configurations (blue) versus a new set of 3600 samples satisfying the Doubly Lee-Courant conditions (green). The DLC schemes show statistically superior equilibrium polarization.}
    \label{fig:dlc_plim_histogram}
\end{figure}

This stringent condition can be understood using the SU(2) formalism. The spin transport through an arc with spin precession $\Psi_k$ followed by a snake with axis $\varphi_k$ is $\mathcal{M}_k = \mathcal{S}(\varphi_k)\mathcal{A}(\Psi_k)$. The DLC condition requires that the combined transport across two consecutive snakes and the arcs following them (by convention), $\mathcal{M}_{k, k+1} = \mathcal{S}(\varphi_{k+1})\mathcal{A}(\Psi_{k+1})\mathcal{S}(\varphi_k)\mathcal{A}(\Psi_k)$, is equivalent to a pure vertical rotation of $\pi$ (or $-\pi$). To have an energy-independent spin phase-advance, all arcs must have the same precession $\Psi$. Together, these constraints simplify to a simple arithmetic relationship between consecutive snake axes:
\begin{equation}
    \varphi_{k+1} - \varphi_k = \pm\frac{\pi}{2} \pmod{2\pi}.
\end{equation}

This condition imposes $2N-1$ constraints on the $2N$ snake axes, leaving one continuous and one discrete degree of freedom: an overall rotation of the entire snake axis pattern, and an overall reflection of all the snake axes. For the HSR ($2N=6$), one such solution are the alternating $\pm45^\circ$ snake axes. The general solution will be an alternating pair of axes separated by \(90^\circ\). To our knowledge, this is the first documented investigation of systematically exploiting this highly symmetric free parameter of snake configurations.

\begin{figure}[H]
    \centering
    \includegraphics[width=1.1\linewidth]{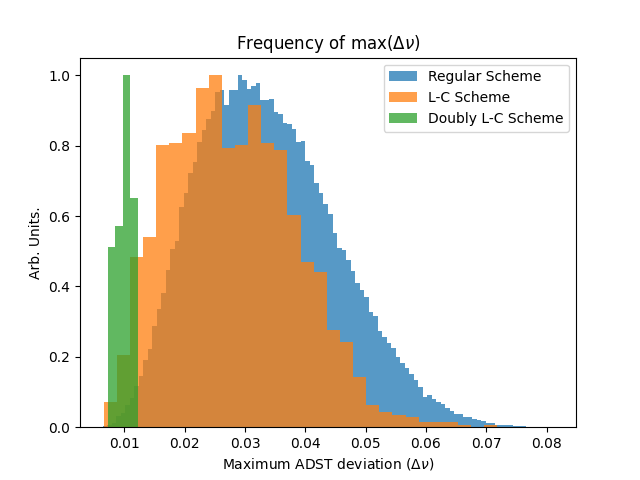}
    \caption{Histogram of \(\max(\Delta\nu)\) for a random sampling of general snake axis configurations (blue) versus the subset satisfying the L-C constraints (orange), versus a new set specifically satisfying the Doubly Lee-Courant conditions (green). Better performance is indicated by smaller ADST spread \(\Delta\nu\).}
    \label{fig:adst_comp}
\end{figure}

Similarly comparing the distribution of \(\min(P_\text{lim})\) for a refined DLC set of snake configurations with the previous general set of solutions show that DLC schemes yield a categorically stronger statistical preference for high polarization, as seen in ~\ref{fig:dlc_plim_histogram}. On the other hand, if we evaluate the distribution of maximum ADST spread max\((\Delta\nu)\) for all three schemes of snake configurations, we again see the success hierarchy of the regular snakes $<$ L-C $<$ DLC in Figure~\ref{fig:adst_comp}.

\begin{table}[H]
\centering
\caption{Snake Axis Schemes for a 2N-Snake Ring}
\label{tab:schemes}
\begin{tabular}{|l|c|c|}
\hline
\textbf{Scheme} & \textbf{Constraints} & \textbf{Free Parameters}\\ 
\hline
General & 1 & $2N-1$ \\
Lee-Courant (L-C) & $N$ & $N$ \\
Doubly L-C (DLC) & $2N-1$ & 1 \\ 
\hline
\end{tabular}
\end{table}

\subsection{Performance Results}
\begin{figure}[H]
    \centering
    \includegraphics[width=\linewidth]{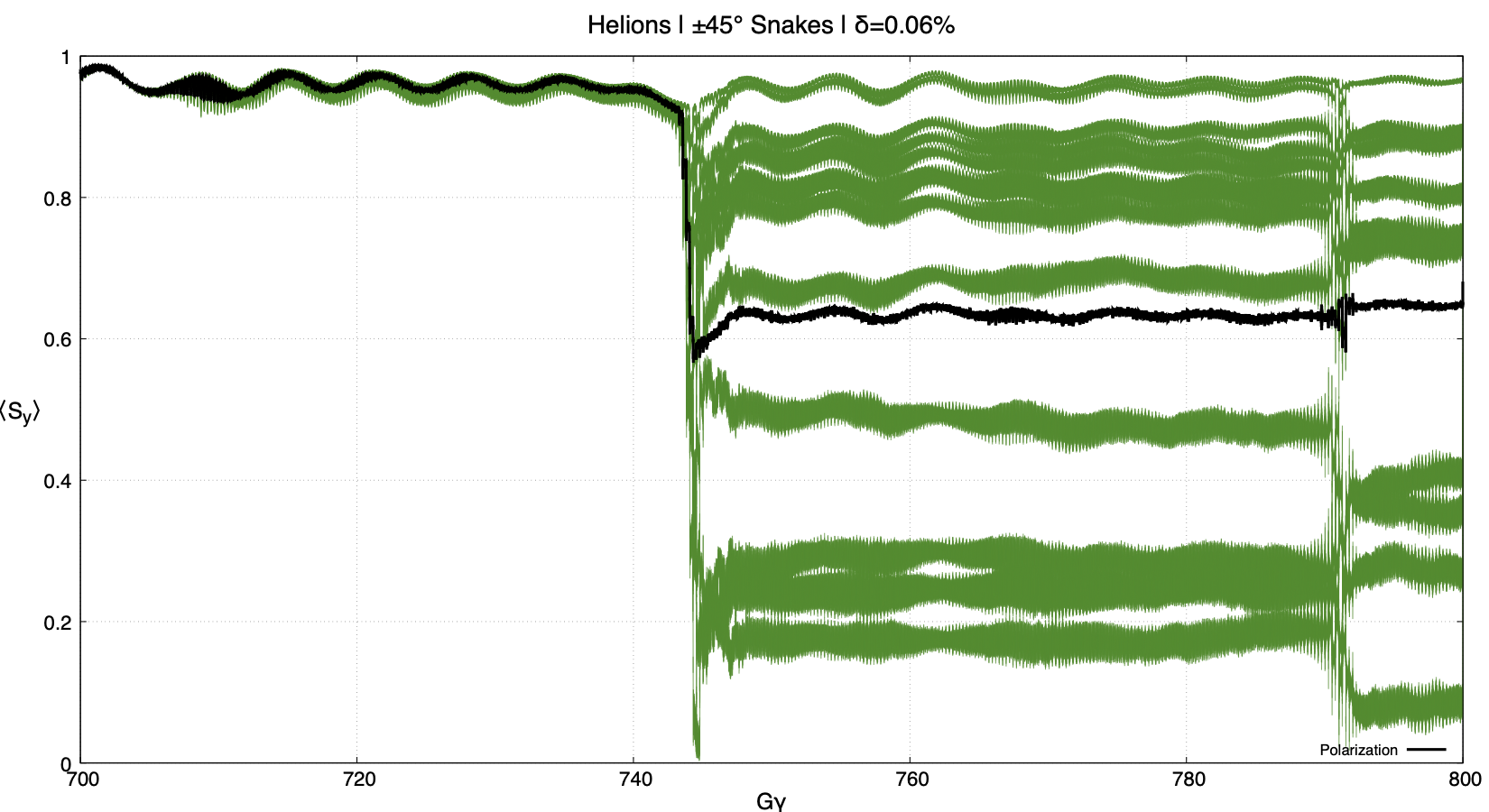}
    \includegraphics[width=\linewidth]{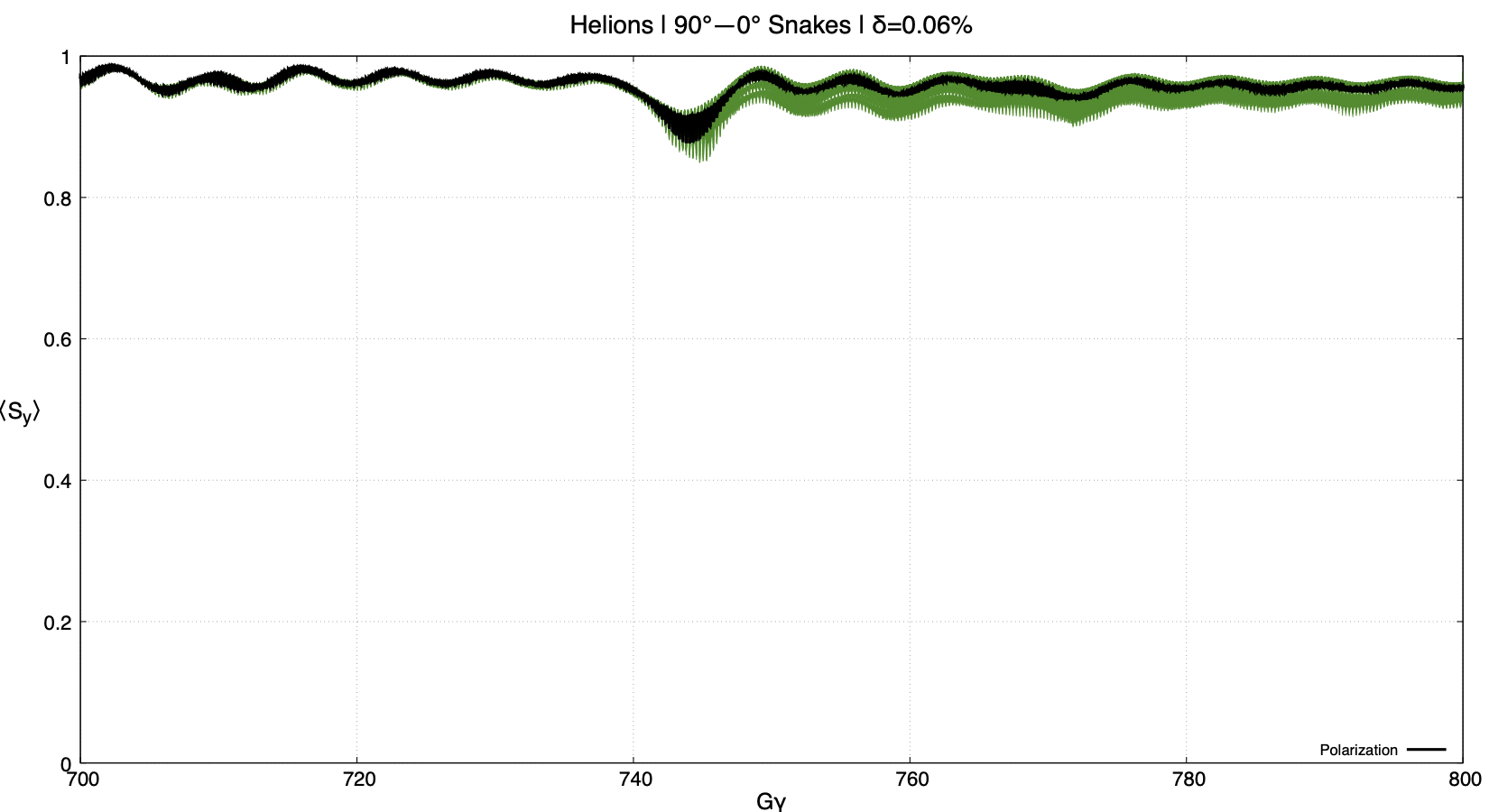}
    \caption{Comparative spin tracking results for the usual choice of $\pm45^\circ$ DLC snake axis scheme, and the DLC scheme with maximal $P_\text{dyn}$. The optimized DLC configuration provides near-perfect polarization transmission.}
    \label{fig:dlc_track}
\end{figure}

The superior performance of the optimal DLC scheme is confirmed with non-perturbative tracking simulations. Figure \ref{fig:dlc_track} compares the polarization survival for the optimal DLC scheme of $90^\circ-0^\circ$ axes and the standard DLC choice of $\pm45^\circ$ axes. The optimal DLC configuration provides exceptionally stable polarization, navigating the entire energy range with minimal loss. The key to this performance is its energy insensitivity. Because the DLC scheme enforces a local cancellation of spin precession that is independent of the arc precession angle ($G\gamma$), it is inherently robust over a wide energy range, unlike other schemes that rely on global, energy-dependent cancellations.

\section{Comparative Analysis and Discussion}\label{sec:6}

A comparative analysis of the two strategies, betatron phase optimization (Method I) and snake axis optimization (Method II), reveals their distinct strengths and underlying philosophies. While a direct quantitative comparison of the final polarization values is not appropriate, as the optimizations were performed on different lattice models and with different beam parameters, the qualitative nature of the solutions achieved by each method can be clearly contrasted. Method I, the optimization of betatron phase, demonstrates that polarization can be preserved by meticulously arranging the orbital dynamics to force a cancellation of depolarizing effects. It is a powerful corrective technique. In contrast, Method II, the optimization of snake axes, addresses the problem at a lower order level because it addresses closed-orbit behavior rather than linear and nonlinear optics. By imposing a high degree of local symmetry on the spin transport, particularly with the DLC scheme, arcs cancel spin depolarization more often among themselves. 

\paragraph{Robustness:} Method I is a form of \textit{numerical compensation}. It relies on fine-tuning the orbital phase to cancel depolarizing effects that are still fundamentally present. This makes the solution inherently energy-dependent and potentially sensitive to machine errors that can disrupt the precise phase relationships. Method II, particularly the DLC scheme, is a \textit{symmetry-driven} solution. It modifies the fundamental spin transport to be locally antisymmetric at a more granular level between the arcs. This approach is intrinsically more robust against variations in energy and is predicted to be less sensitive to lattice imperfections. Sensitivity analysis remains a goal for future studies.

\paragraph{Practicality:} The two methods present a trade-off between operational complexity and hardware requirements. Method I primarily uses existing hardware (quadrupoles). However, it requires precise, and potentially dynamic, control over many magnet circuits during the acceleration ramp, which poses significant operational challenges for control and commissioning, largest of which is the difficulty of rapidly ramping quadrupoles over short periods of time. Method II offers a ``set-and-forget" solution that is robust over the whole energy range. The primary challenge is that ideal DLC configurations often require snake axes far from the longitudinal direction. This leads to larger orbit excursions within the snake magnets, which must be accommodated by the beam aperture and may require specialized snake designs.

\paragraph{Synthesis:} The two methods are not mutually exclusive but are, in fact, complementary. The most effective strategy for the EIC HSR would be a hybrid approach. The DLC scheme should be implemented as a robust, symmetric baseline, which will passively eliminate the vast majority of the resonance strength. Then, the betatron phase advance matching can be used as a fine-tuning tool, applying small, static corrections to compensate for any residual effects from lattice imperfections or the symmetry-breaking IRs.

\section{Conclusion}\label{sec:7}

The scientific mandate of the Electron-Ion Collider requires the acceleration of highly polarized hadron beams to unprecedented energies in a machine whose lattice symmetry is fundamentally broken by its complex interaction regions. This broken symmetry excites a dense spectrum of higher-order spin resonances, presenting a formidable challenge to polarization preservation.

In this paper, we have systematically investigated and compared two distinct strategies for enforcing a ``snake matched" condition in the HSR. The first, optimization of the betatron phase advance, is an effective optical compensation technique. We demonstrated that by carefully tuning the lattice quadrupoles in an energy-dependent manner, or by finding a robust fixed-optics compromise, it is possible to navigate the strongest resonant regions and maintain a high degree of polarization. However, this method relies on a delicate numerical cancellation and is fundamentally energy-dependent.

The second strategy, optimization of the snake rotation axes, offers a more powerful and fundamental solution by manipulating the closed-orbit spin dynamics. Our analysis revealed a clear hierarchy of performance based on the degree of imposed symmetry. By introducing and formalizing the \emph{Doubly Lee-Courant (DLC) scheme}, a novel class of configurations with maximal local spin symmetry, we have identified a solution that provides exceptional and energy-robust polarization preservation. The DLC principle of enforcing a $\pi$ spin phase advance across every consecutive pair of snakes suppresses the underlying resonance drivers at their source.

This paper contains the first identification and optimization of the only free parameter in DLC schemes. This represents a 1D subset of the large $2N-1$ dimensional parameter space of snake axes for the design of high-energy polarized beam facilities. We conclude that the optimal path forward for the EIC HSR, and for future projects with odd super-periodicity, may be a hybrid approach: employ a highly symmetric snake configuration like the DLC as a robust baseline to make the ring inherently more polarization-secure, and use betatron phase adjustments for fine-tuning corrections of residual imperfections. The symmetry principles uncovered in this study provide a flexible and powerful new toolkit for ensuring the delivery of high-polarization beams, a critical requirement for the successful realization of the EIC's scientific program.

\section*{Acknowledgments}
The authors would like to thank Joseph Devlin and Desmond Barber for many insightful discussions and collaborations on this topic, and David Sagan for his tireless efforts in building and supporting the Bmad toolkit. Furthermore, we would like to thank Vincent Schoefer, Haixin Huang, Kiel Hock, Vadim Ptitsyn, and Vahid Ranjbar for many fruitful discussions about polarization transmission and J. Scott Berg and Henry Lovelace III for helpful discussions on lattice implementation. Finally, we would like to collectively acknowledge the entire Brookhaven team for the many years of successful polarized proton operation at RHIC.

This work has been supported by Brookhaven Science Associates, LLC under Contracts No. DE-SC0012704 and No. DE-SC0018008 with the U.S. Department of Energy.

\bibliography{ref}

\end{document}